\documentclass[9pt]{sig-alternate}
\usepackage{color,soul}
\usepackage{psfrag}
\usepackage{algorithmic}
\usepackage{amsmath}

\usepackage[ruled,vlined]{algorithm2e}
\usepackage{pgfplots}
\usepackage{subfigure}
\usepackage{multirow}
\usepackage{url}
\usepackage{balance}
\usepackage[margin=0pt,skip=0pt,font={bf,small},labelfont=bf,aboveskip=5pt,belowskip=-10pt]{caption}

\newcommand{\squishlist}{
   \begin{list}{$\bullet$}
    { \setlength{\itemsep}{0pt}      \setlength{\parsep}{3pt}
      \setlength{\topsep}{3pt}       \setlength{\partopsep}{0pt}
      \setlength{\leftmargin}{1.0em} \setlength{\labelwidth}{1em}
      \setlength{\labelsep}{0.5em} } }

\newcommand{\squishend}{
    \end{list}  }

\begin{document}

\title{
SARA: Self-Aware Resource Allocation \\
for Heterogeneous MPSoCs
}

\numberofauthors{1}
\author{\alignauthor Yang Song{$^{\dag}$}, Olivier Alavoine{$^{\ddag}$}, Bill Lin{$^{\dag}$}\\
\affaddr{$^{\dag}$Electrical and Computer Engineering Department, University of California at San Diego} \\
\affaddr{$^{\ddag}$Qualcomm Inc., San Diego, CA}\\
\affaddr{y6song@ucsd.edu}}

\maketitle

\begin{abstract}
In modern heterogeneous MPSoCs, the management of shared memory resources is crucial in delivering end-to-end QoS.
Previous frameworks have either focused on singular QoS targets or the allocation of partitionable resources among CPU applications at relatively slow timescales. However, heterogeneous MPSoCs typically require instant response from the memory system where most resources cannot be partitioned. Moreover, the health of different cores in a heterogeneous MPSoC is often measured by diverse performance objectives.
In this work, we propose a Self-Aware Resource Allocation (SARA) framework for heterogeneous MPSoCs. Priority-based adaptation allows cores to use different target performance and self-monitor their own intrinsic health. In response, the system allocates non-partitionable resources based on priorities. The proposed framework meets a diverse range of QoS
demands from heterogeneous cores.
\end{abstract}


\section{Introduction}\label{sec_introduction}

\begin{figure}[!htb]
\centering
\includegraphics[width=0.35\textwidth]{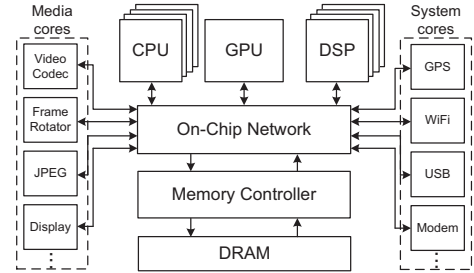}
\caption{Heterogeneous system architecture example.}
\label{hsa}
\end{figure}

Modern heterogeneous MPSoCs \cite{snapdragon,tegra} have been widely deployed in mobile devices thanks to their energy efficiency. These MPSoCs typically integrate a diverse collection of cores.
Fig.~\ref{hsa} depicts an example of a heterogeneous MPSoC.
Besides general-purpose cores like the CPU for running applications, most heterogeneous cores are dedicated to certain functions, such as the GPU, the DSP and the display. These cores have diverse notions of Quality-of-Service (QoS). For example, the GPU measures target real-time performance in terms of frame rate; the DSP demands the memory latency to remain below a certain limit; and the display requires sufficient bandwidth to refresh frames at a constant rate.

\begin{figure}[!htb]
\centering
\includegraphics[width=0.5\textwidth]{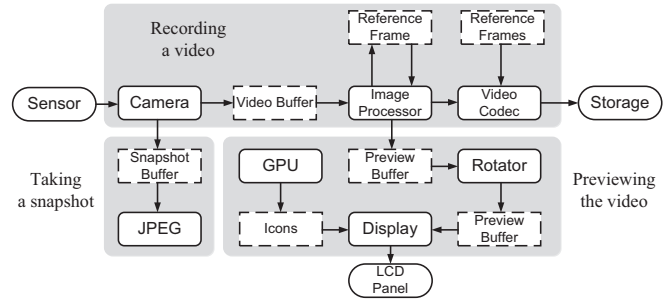}
\caption{Simplified dataflow of a camcorder application. Shared memory represented by boxes in dash lines and cores by boxes in solid lines.}
\label{data_flow}
\end{figure}

To save cost and energy, heterogeneous cores commonly share resources, among which, the sharing of the memory system (including the on-chip network and the memory controller) is the most challenging because memory performance often has a direct and substantial impact on the system performance.
As data is being shared through memory, competing memory requests from different cores interfere with each other, and these memory interferences can cause the memory system to fail in meeting the target performance of some cores.
Fig.~\ref{data_flow} depicts a camcorder application, which represents a typical use case
in that it involves many cores at the same time. With ineffective memory scheduling, a real-time core (e.g., the display) may not achieve the target real-time performance due to inadequate memory bandwidth. Moreover, as latency-sensitive cores such as the DSP
share memory with other cores, they can be easily overwhelmed by real-time cores consuming high bandwidth.

QoS-aware management for specific types of memory resources has been well-studied by previous work \cite{austin,cmu,stvq,pvc,cache_dac16}. In \cite{austin}, a QoS-aware scheduling policy was proposed for CPU-GPU systems. The concept of frame progress was introduced for monitoring GPU performance. Although the policy can be extended to include more media cores, it cannot be applied to real-time cores whose target QoS cannot be assessed in terms of frame rate.
Moreover, holistic memory management frameworks for CPU-centric homogeneous systems have also been explored recently \cite{mete,seec_dac,arcc}. This series of work typically constructs a management model based on the control theory to partition computing and memory resources. These frameworks accept flexible QoS targets as clients are allowed to define their own target performance. Nonetheless, such type of approaches is performed at a relatively slow timescale (e.g., on the order of milleseconds) due to the computational complexity. In comparison, real-time cores in heterogeneous MPSoCs often demand much more instant response from the memory system.
Besides, communication between heterogeneous cores is mainly conducted through shared memory as shown in Fig.~\ref{data_flow}, because multimedia data is generally too large to fit in caches. Therefore, DRAM plays a more crucial role in heterogeneous systems. However, previous frameworks cannot handle DRAM effectively because its bandwidth is not partitionable. Specifically, available DRAM bandwidth relies on the memory access pattern, as higher spatial locality results in fewer redundant precharge operations and better memory efficiency.

So far, there has not been a QoS-aware resource management model for heterogeneous MPSoCs which is capable of allocating non-partitionable resources to fleeting QoS demands.
In this work, we propose the Self-Aware Resource Allocation (SARA) framework as a solution.
The contributions of our work can be summarized as follows.
\squishlist
\item We propose a QoS-aware holistic resource management framework for heterogeneous systems. The SARA model accepts diverse notions of QoS and monitors performance distributively with lightweight meters to guarantee end-to-end QoS.

\item We introduce priority-based self-adaptations for the management of non-partitionable resources, such as DRAM and on-chip network,
which constitute most of the shared resources in heterogeneous MPSoCs.

\item We evaluate the proposed framework using memory traffic of next-generation MPSoCs and show that the proposed SARA model delivers target performance to all cores. In contrast, the performance of critical cores can fall below 10\% of their targets without the SARA framework. Further, memory system optimization is performed without QoS degradations.
\squishend


The rest of this paper is organized as follows:
Section \ref{sec_related} briefly reviews related work.
Section \ref{sec_systemmodel} describes the proposed SARA framework.
Experimental results and conclusions follow in Sections \ref{sec_experiment} and \ref{sec_conclusions}.

\section{Related Work}\label{sec_related}
Most previous work on QoS-aware resource management in heterogeneous MPSoCs were focused on a single layer of the memory system.
In \cite{austin}, a novel scheduling policy was introduced to dynamically balance bandwidth between the CPU and the GPU based on the frame progress of real-time workloads. To achieve QoS-aware memory scheduling, the staged memory scheduler \cite{cmu} was presented as the first QoS-aware scheduler for CPU-GPU systems. Further, the single-tier virtual queuing memory controller \cite{stvq} was proposed to overcome the limitation of two-tier schedulers in QoS-aware scheduling. Besides memory scheduling, QoS-aware cache management \cite{cache_dac16} and on-chip network design \cite{pvc} have also been well-explored in recent years. Nonetheless, these work cannot guarantee end-to-end QoS because they only deal with certain parts of the memory system. For example, the QoS provided in the memory controller could be deteriorated by the interconnect if it is not applying the same QoS policy.
In addition, implementing a centralized QoS monitor in the memory system can be prohibitive since it needs to collect runtime information from all cores.
More limiting, these work assume specific notions of QoS, which is not applicable to modern heterogeneous MPSoCs where the health of different cores is often evaluated by diverse performance objectives.

METE \cite{mete} is a multi-level framework for end-to-end resource management based on the control theory. It utilizes runtime information to predict application behaviors. Application controllers calculate the amounts of resources required to achieve target application performance. A global resource broker determines the final resource partitions for applications.
SEEC \cite{seec_dac} is a self-aware computing framework designed for a many-core processor. It follows the control loop of observe-decide-act for resource allocations. Performance of CPU applications are observed by the decision engine which decides resource partitions using available actions defined by system designers.
ARCC \cite{arcc} is a self-computing framework implemented in the Tessellation many-core OS. It performs the two-level scheduling: first the resource allocation broker distributes global resources and then at user-level scheduling policies are customized separately.

Aforementioned frameworks were intended for CPU-centric multi-core systems. These frameworks are aimed at allocating partitionable resources, such as CPU cores and cache ways, to applications at the software/OS level. They are not suitable for heterogeneous MPSoCs for the following reasons.
First, complicated control models may not be fast enough for heterogeneous cores
(e.g., these software/OS level approaches operate at milleseconds timescales).
For example, the DSP sets limit on memory latency at nanosecond level, but prior frameworks need more time to adapt through control theory computations in OS.
Second, prior work assume all memory resources are partitionable. However, DRAM bandwidth cannot be simply partitioned like cache ways.
In DRAM, data storage of a memory bank is organized into rows and columns.
To access a column, the row where this column is located will be loaded into the row-buffer (i.e. row activation operation) after the other rows are closed (i.e. precharge operation) \cite{memory_systems}.
These row activation and precharge operations cause time penalty without contributing to actual data transfer, which makes DRAM bandwidth inconstant and unpredictable.



\section{Self-Aware Resource \\ Allocation Framework}\label{sec_systemmodel}


\begin{figure}[!htb]
\centering
\includegraphics[width=0.45\textwidth]{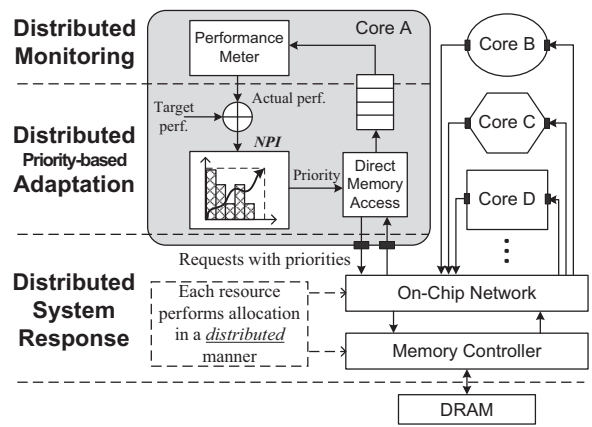}
\caption{The proposed SARA framework for heterogeneous MPSoCs. Each core self-monitors its performance and self-adapts its priority, and each resource performs priority-based allocation in a distributed manner.}
\label{sara_model}
\end{figure}

\begin{figure}[!htb]
\centering
\includegraphics[width=0.5\textwidth]{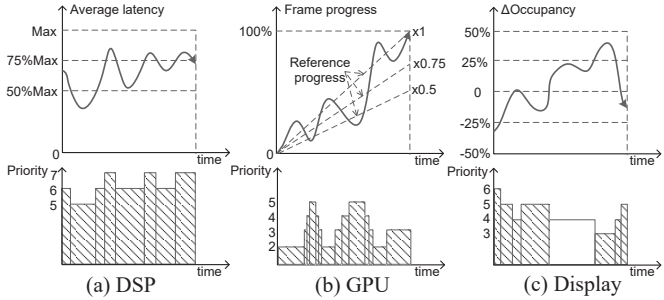}
\caption{Examples of priority-based adaptation in heterogeneous cores,
including the DSP, GPU and display.}
\label{adaptation}
\vspace{-2mm}
\end{figure}


The proposed architecture of SARA framework is shown in Fig.~\ref{sara_model}. The resource management model consists of three stages, including distributed monitoring, priority-based runtime adaptation and system response. In the rest of this section, we will go through SARA framework stage by stage.

\subsection{Distributed Self-Monitoring}

In the first stage, each core self-monitors its own performance.
The distributed monitoring relieves the memory system from the burden of monitoring heterogeneous cores with various notions of QoS.
Self-monitoring also provides more accurate feedback on the end-to-end QoS compared with centralized monitoring in the memory system.
In addition, implementing lightweight performance meters is good for scalability, because a new core can be added or modified without updating the rest of the system.

Every core customizes its own internal performance meter to measure its own performance or progression against a given target, and
the measurement gets normalized
into a fractional number
called a Normalized Performance Indicator (NPI),
which is used as
an indicator of
the core's intrinsic health.
In the DSP, the performance meter monitors the average latency of its transactions, while in the display the meter counts the occupancy level in the read buffer.
The
deviation
from the target performance (e.g., latency, occupancy level, etc) produces the NPI metric.
In our framework, each independent DMA (Direct Memory Access) unit is equipped with a performance meter.
Note that there are usually multiple DMAs in a single core. For simplicity, we only show one DMA per core in Fig.~\ref{sara_model}.


\subsection{Distributed Priority-Based Adaptation}\label{sec_adaptation}
In the second stage, each core adapts the relative priority of its transactions based on its NPI value.
The NPI value delivered by the performance meter is translated into a relative priority level which is attached to memory transactions from the same DMA. The priority level will be evaluated within
on-chip network
arbiters and the memory controller, as the transaction travels
along the way to DRAM. 
Priority-based arbitrations allow the memory system to provide QoS without specifying the heterogeneous QoS for all cores and DMAs.
Same with performance meters, the
formulation of the NPI metric and the adaptations of priority can be implemented differently from core to core, depending on the local target performance. Fig.~\ref{adaptation} shows three examples of priority-based adaptation in different cores.

As for the DSP, the target performance is to have the average memory latency lower than the maximum latency limit. The average latency is measured and compared with a pre-set limit to produce the NPI value (see Eqn. \ref{kpi_dsp}), which remains above or equal to 1 when the target performance is achieved.
This NPI value is then translated to a relative priority level (Fig.~\ref{adaptation}(a)). The priority level increases along with average latency.
\begin{equation}\label{kpi_dsp}
\scriptsize
NPI_{DSP} = \frac{maximum \; latency \; limit}{average \; latency}
\end{equation}

Similarly, cores requesting for bandwidth produce NPI metrics by computing the ratio between the average and the target bandwidth. However, frame rate differs from bandwidth, because frame size can be variable and thus a constant frame rate can lead to variable bandwidth.
Hence frame progress \cite{austin,stvq} is used instead to produce NPI metrics for frame rate based cores.
Take the GPU as an example,
the target is to let the frame progress reach 100\% as the current frame period comes to an end. The GPU's NPI value is produced at any time by comparing the frame progress with reference progresses which grow proportionally with frame time.
The NPI value is then translated to a relative priority level of GPU transactions.
Fig.~\ref{adaptation}(b) shows the reference progresses achieving 1, 0.75 and 0.5 times the average data rate of target performance.
\begin{equation}\label{kpi_gpu}
\scriptsize
NPI_{GPU} = \frac{frame \; progress}{reference \; progress}
\end{equation}

In the display, LCD panel reads data from a read buffer at a constant frame rate, while the display controller DMA tries to refill this buffer from DRAM so it never gets empty. Its \emph{health} (see Eqn. \ref{kpi_display}) relies on maintaining the refill rate ($R_{refill}$) no lower than the read data rate ($R_{read}$), and can be indicated by the variation of buffer occupancy level ($\Delta occupancy$).
Compared with an initial level (e.g. 50\%), the lower the occupancy level of this buffer gets, the worse the NPI value becomes, which is in turn translated to a higher priority level (Fig.~\ref{adaptation}(c)).
\begin{equation}\label{kpi_display}
\scriptsize
NPI_{display} = \frac{R_{refill}}{R_{read}} = 1 + \frac{\Delta occupancy}{R_{read} \cdot time}
\end{equation}

Intuitively, one might be concerned that every core would intentionally raise the priority to the maximum level to obtain as
much resources as possible. However, this situation should not happen because the priority level is only maximized when the actual performance is far
below
the target.
The system designer has the responsibility to make sure cores have realistic
performance targets
and
enough resources to satisfy all possible combinations of QoS demands.
Once the system is fabricated in hardware, heterogeneous cores cannot change their target performance arbitrarily, especially because most of them are fixed-function IP blocks with invariable QoS targets and little programmability.




In our evaluations, the priority levels are quantized into $2^k$ levels, which can be encoded using $k$ bits.  We found that $k = 3$ bits provides sufficient granularity in priority levels to produce satisfying results (i.e., the priority levels range from 0 to 7).



\subsection{Distributed System Response}\label{sec_system_response}

As transactions travel through the memory system, the system responds to QoS demands by providing resource management based on their priority levels.
The priority-based management is performed correspondingly in different parts of the memory system.
In on-chip network routers, transactions with higher priorities are preferentially selected during switch allocation.
In the memory controller, when a priority-based scheduler arbitrates among transactions going to available memory banks, the ones with higher priorities have more chances to be served. An example of such memory scheduling policies is the priority-based round-robin shown in Policy 1.
To avoid starvation of transactions with low priorities, the scheduler also needs to consider the aging factor during arbitration. In our evaluations, the scheduler periodically clears the backlog of transactions that have waited for at least $T$ cycles (e.g., $T = 10000$ cycles).

\squishlist
\item \textbf{Policy 1}: Suppose $P_A$ and $P_B$ are priorities for transactions A and B, if $P_A > P_B$ choose A; if $P_A < P_B$ choose B; otherwise choose between A and B in round-robin manners.
\squishend

Priorities notify the system whether the cores are in urgent QoS demands.
That gives the memory system an opportunity to optimize memory performance without undermining the QoS.
Specifically, when transactions are in low urgency,
the system can improve memory performance such as row-buffer hit rate, instead of focusing on serving QoS demands.

Row-buffer hits refer to the number of memory accesses to the same active row-buffer before precharge. More row-buffer hits means less time and power are wasted on row activation and precharge operations.
Thus increasing row-buffer hits helps lower memory latency and improve DRAM total bandwidth.

To increase row-buffer hits, the memory controller re-orders transactions to favor the ones hitting open rows.
It may cause degradations to the QoS when the transactions in high urgency are postponed due to row-buffer hits optimization.
Yet, with priorities, the memory controller is aware of the urgency levels of transactions and able to avoid delaying urgent transactions during optimization.
Policy 2 shows an extension of Policy 1 to increase row-buffer hits without QoS degradations.
The parameter $\delta$ is an adjustable threshold to balance row-buffer hits optimization and QoS-aware scheduling.
When the priority level is lower than $\delta$, the scheduler focuses on row-buffer hits, otherwise the QoS comes first.
A higher $\delta$ value gives more favor to DRAM bandwidth, but also potentially causes more disturbance to the QoS.
We found $\delta = 6$ a good setting to achieve high DRAM bandwidth without causing QoS degradations.

\squishlist
\item \textbf{Policy 2}: Suppose transaction A is going to an active row-buffer and B is not. If $P_A, P_B < \delta $ or $P_A = P_B$, choose A. Otherwise, perform priority-based round-robin.
\squishend

The priority-based resource allocation is able to handle non-partitionable with little computation in comparison with previous management models \cite{mete,seec_dac,arcc}. This facilitates instant response from the memory system to QoS demands.



\subsection{Hardware Implementation}
The implementation of the proposed SARA framework includes three parts: the computation of NPI value, the translation of NPI value to a priority level, and the priority-based arbitration in the memory system.

To calculate the NPI, a divider is needed at the performance meter for each DMA.
For the translation of the NPI, a mapping function can be stored in a look-up table at each core. Each priority level is assigned with a table entry, and this entry stores the lowest NPI value allowed at that priority level. For example, if priority $= p$ when \emph{NPI} $ \in [u, v)$, the value $u$ will be stored at the entry for $p$ on the look-up table. Note that $v$ will be the lower bound of the NPI for the priority level $p-1$.
Comparators are needed to access table entries in parallel. If the current NPI value is not lower than the stored lower bound of NPI value, the corresponding priority level will be asserted.
When multiple priority levels are asserted, the lowest level will be adopted.

Supposed each priority level is encoded into three bits, 
a look-up table requires $2^3 = 8$ entries and each entry is a register for the NPI value. A 
comparator is paired with each table entry.
In total, the implementation only costs the storage of eight registers 
and eight 
comparators per core.

In the memory system, performing the priority-based arbitration requires a 3-bit comparator to arbitrate among transactions with different priority levels. Since most existing QoS-aware schedulers already provide hardware support for priorities, our framework can be integrated into the memory system without raising complexity.

\section{Evaluation}\label{sec_experiment}

\begin{table}
\centering
\scriptsize
    \caption{Simulation settings.}
    \label{setting}
    \begin{tabular}{c|c}
    \hline
    \hline
        \multicolumn{2}{c}{Test Cases} \\
    \hline
         \multirow{2}{0.4in}{Case A} & all cores active \\
         & with DRAM @ 1866MHz; \\
    \hline
         \multirow{3}{0.4in}{Case B} & inactive cores: \\
         & GPS, camera, rotator and JPEG, \\
         & with DRAM @ 1700MHz. \\
    \hline
    \hline
        \multicolumn{2}{c}{Memory Controller} \\
    \hline
         Total entries & 42 \\
    \hline
         Transaction queues & 5\\
    \hline
    \hline
         \multicolumn{2}{c}{DRAM} \\
    \hline
        Volume  & 2GB \\
    \hline
        Max I/O bus freq. & 1866MHz \\
    \hline
        CL-tRCD-tRP (cycles) & 36-34-34 \\
    \hline
        tWTR-tRTP-tWR (cycles) & 19-14-34\\
    \hline
        tRRD-tFAW (cycles) & 19-75 \\
    \hline
        Channels-Ranks-Banks & 2-2-8 \\
    \hline
    \hline
    \end{tabular}
\end{table}

In this section, the proposed SARA framework will be tested to demonstrate its effectiveness in providing target performance to heterogeneous cores. Two test cases based on the camcorder dataflow (Fig.~\ref{data_flow}) will be used for demonstration. Further, we will show row-buffer hits optimization can be performed efficiently within SARA framework without performance degradations.

The proposed framework is modeled as in Fig.~\ref{sara_model}, where memory traffic from every DMA is generated based on a next-generation MPSoC \cite{snapdragon}. DRAMSim2 \cite{dramsim} with LPDDR4 timing model is used for cycle-accurate simulation of DRAM.
Table \ref{setting} shows the simulation settings.
Table \ref{qos_table} lists the simulated cores and the types of target performance.

The target performance for each core is set according to the camcorder dataflow (Fig.~\ref{data_flow}) which runs at 30fps.
For instance, the frame rotator writes and reads 1080p YUV420 images at 30fps, which requires 89MB/s for each DMA and 178MB/s in total.

\begin{table}
\vspace*{4mm}
\centering
\scriptsize
    \caption{Summary of heterogeneous cores and types of target performance.}
    \label{qos_table}
    \begin{tabular}{c|c||c|c}
    \hline
        \multirow{2}{*}{Name} & Performance  & \multirow{2}{*}{Name} & Performance \\
                              & type         &                       & type \\
    \hline
        GPU & frame rate
& Display & buffer occupancy \\
        DSP & latency
& GPS & processing time \\
        Image Processor & frame rate
& WiFi & bandwidth \\
        Video Codec & frame rate
& USB & bandwidth \\
        Rotator & frame rate
& Modem & processing time \\
        JPEG & frame rate
& Audio & latency  \\
	Camera & buffer occupancy
&	&	\\
    \hline
    \end{tabular}
\vspace*{-4mm}
\end{table}

\subsection{Delivering Target Performance}


\begin{figure*}[!htb]
\centering
\scriptsize
\includegraphics[width=1\textwidth]{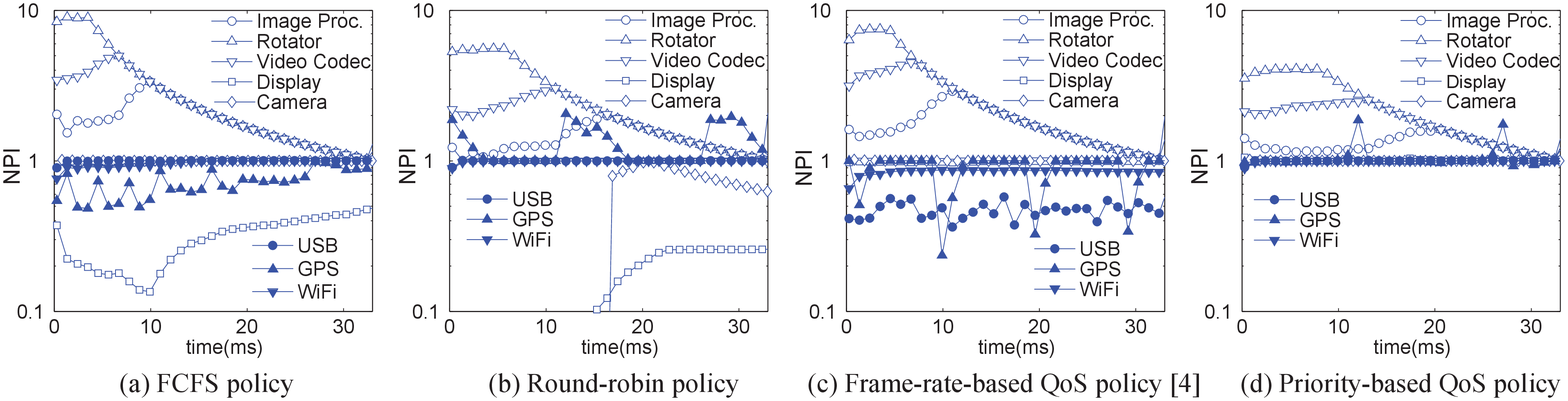}
\caption{NPI value of critical cores during one frame period (33ms) for test case A with different arbitration policies.}
\label{kpi_case_a}
\end{figure*}


\begin{figure*}[!htb]
\centering
\scriptsize
\includegraphics[width=1\textwidth]{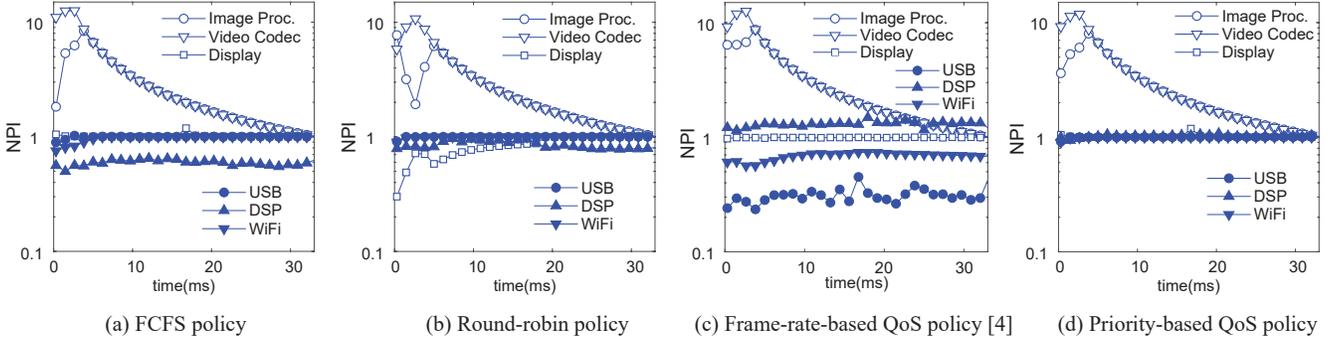}
\caption{NPI value of critical cores during one frame period (33ms) for test case B with different arbitration policies.}
\label{kpi_case_b}
\end{figure*}

To begin with, we test the SARA framework in delivering target performance to heterogeneous cores.
For comparison, four arbitration policies are used in the memory controller and on-chip network arbiters, including first-come-first-serve (FCFS), round-robin (RR), a frame-rate-based QoS policy \cite{austin} and the priority-based QoS policy (Policy 1). FCFS policy serves all the transactions according to the arrival order. Round-robin policy separates transactions into different queues and serves them in a round-robin fashion. In the memory controller, we have five transaction queues respectively designated to the CPU, the GPU, the DSP, media cores and system cores. Round-robin policy also applies to on-chip network arbiters, as input queues are served in turn.
The frame-rate-based QoS policy prioritizes media cores when they are missing real-time deadlines, but otherwise, the policy provides best-effort service to latency-sensitive cores.
Furthermore, the priority-based QoS policy compares priority levels for arbitration and uses round-robin as the tiebreaker.

The NPI of critical cores during a frame period are shown in Fig.~\ref{kpi_case_a} when test case A is applied.
As explained in Section \ref{sec_adaptation}, the NPI metric reflects performance as higher value indicates better performance. When NPI value drops below 1, it means the the target performance is not achieved.

Without reordering memory requests, FCFS policy ends up spending most of the time serving cores consuming high bandwidth. That easily leads to the starvation of latency-sensitive cores. As shown in Fig.~\ref{kpi_case_a}(a), the NPI of the GPS drops below 1 because the GPS is overwhelmed by other system cores sharing the same interconnect, such as the USB. For media cores, the video codec, the rotator and the image processor have all the frame data available at the beginning of a frame period and thus create bursty traffic, meanwhile the camera and the display generate and consume data at constant rates which are determined by image sensor and LCD panel. In Fig.~\ref{kpi_case_a}(a), media cores with bursty traffic obtain most of the bandwidth in the beginning, resulting in high NPI value. On the other hand, the display fails to achieve the target performance.
The display's NPI drops as low as 0.13 which means only 13\% of the target performance is achieved.

When round-robin policy is applied, the competition among media cores becomes more intense since they share the same transaction queue in the memory controller. In Fig.~\ref{kpi_case_a}(b), the display and the camera both fail due to the interference from other media cores. Less than 10\% of their target performance is achieved in the worst case. In the meantime, all the system cores meet their target performance because they avoid the interference from media cores by using a separate transaction queue.

The frame-rate-based QoS policy helps all media cores achieve NPI value above 1 in Fig.~\ref{kpi_case_a}(c). However, all system cores fail due to the absence of adaptations for the cores with different QoS targets other than frame rates.

In Fig.~\ref{kpi_case_a}(d), all the cores reach their target performance when QoS-aware scheduling is performed, because priority-based adaptations help arbiters serve the cores in urgent needs. 
Note that the NPI of the other cores such as the GPU are not shown because no failure is observed from these cores.

The results by test case B are shown in Fig.~\ref{kpi_case_b}. Similar to Fig.~\ref{kpi_case_a}, the latency-sensitive DSP suffers when FCFS policy is adopted (Fig.~\ref{kpi_case_b}(a)). When round-robin policy is applied (Fig.~\ref{kpi_case_b}(b)), the DSP suffers less since it has its own transaction queue, while the display fails due to the increased interference from other media cores sharing the same transaction queue. Again, the frame-rate-based QoS policy fails to serve non-media cores. At last, the dynamic priorities help the memory system deliver target performance to all cores (Fig.~\ref{kpi_case_b}(d)).

\begin{figure}[!htb]
\centering
\includegraphics[width=0.5\textwidth]{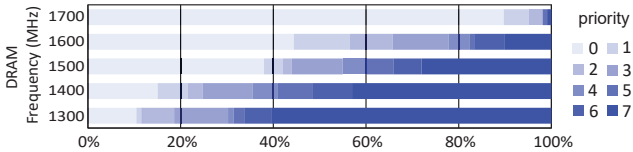}
\caption{Distributions of the image processor's priority levels during one frame period (33ms) with respect to different DRAM frequencies.}
\label{pri_freq}
\end{figure}

Next, we take the image processor from test case A as an example to examine the priority-based adaptation in a single core.
Fig.~\ref{pri_freq} shows the distributions of the image processor's priority levels during one frame period, while DRAM frequency decreases from 1700MHz to 1300MHz.
Each horizontal bar is designated to a certain DRAM frequency.
In a single bar, each block represents the percentage of time during which a certain priority level is adopted.
Different shades of blue represent different priority levels, as higher priority levels in darker shades.
As shown in Fig.~\ref{pri_freq}, when DRAM frequency is set to 1700MHz, for 90\% of the time the image processor is adapted to the priority of 0. As frequency decreases, less memory requests can be processed by DRAM. More memory interferences and competitions happen as the result. To maintain target bandwidth, the self-adaptation leads to a gradual increase in priority levels, which can be observed through the increasing area of blocks in dark shades.
When DRAM frequency is lowered to 1300MHz, the image processor has the priority of 7 for 60\% of the time.
In addition, as frequency decreases, the average bandwidth of the image processor remains above target bandwidth thanks to the priority-based adaptation.


\subsection{Row-buffer Hits Optimization}\label{sec_rb_opt}
As explained in Section \ref{sec_system_response}, row-buffer hits optimization helps improve available DRAM bandwidth.
With the knowledge of heterogeneous cores' urgency levels, the memory controller in the SARA framework is capable of optimizing row-buffer hits without degrading system performance.

\begin{figure}[!htb]
\centering
\includegraphics[width=0.45\textwidth]{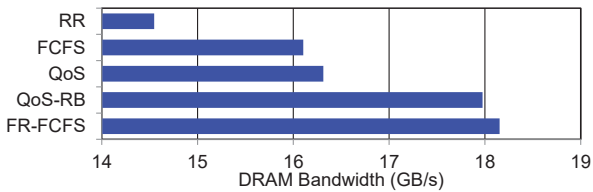}
\caption{Summary of average bandwidth when different scheduling policies applied.}
\label{total_bw}
\end{figure}
\begin{figure}[!htb]
\centering
\includegraphics[width=0.5\textwidth]{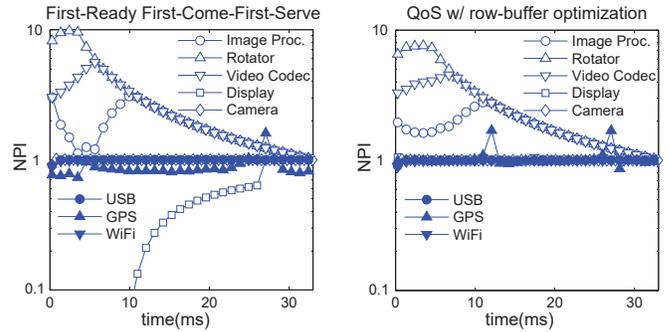}
\caption{NPI value for test case A with respect to FR-FCFS and QoS-RB scheduling policies.}
\label{rb_opt}
\end{figure}

For comparison, we compare with another scheduling policy named first-ready first-come-first-serve (FR-FCFS) which prioritizes transactions going to open rows whenever it is possible, and otherwise schedules transactions based on FCFS. FR-FCFS policy is expected to achieve the most row-buffer hits and the highest DRAM bandwidth.
Fig.~\ref{total_bw} shows the average DRAM bandwidth during one frame period when test case A is applied. Four memory scheduling policies are tested, including RR, FCFS, QoS (Policy 1), QoS-RB (Policy 2) and FR-FCFS. Fig.~\ref{rb_opt} shows the NPI of critical cores as QoS-RB and FR-FCFS are adopted.
As expected, FR-FCFS policy achieves the highest bandwidth, whereas performance degradations happen to the GPS and the display as the expense.
The bandwidth by QoS-RB is slightly lower (by 1\%) than FR-FCFS, but much higher than other policies. Specifically, the average DRAM bandwidth obtained by QoS-RB policy is 24\%, 12\% and 10\% higher than RR, FCFS and QoS policies respectively. In the meantime, no performance degradations are caused to heterogeneous cores.


\section{Conclusions}\label{sec_conclusions}
In this work, we proposed the self-aware resource allocation (SARA) framework for memory management in heterogeneous systems. Lightweight performance meters are distributed in each core to monitor end-to-end QoS with low cost.
The priority-based adaptation allows cores to customize their target performance and adjust their priority levels according to the observed performance.
The memory system with non-partitionable resources responds to QoS demands by performing priority-based management which does not require complicated computations.
Experimental results show that with the priority-based adaptation and management, SARA framework helps all the heterogeneous cores achieve their target performance. By comparison, without using priorities, performance of critical cores can drop lower than 10\% of the target.


\begin{scriptsize}
\bibliographystyle{unsrt}
\bibliography{literature}
\end{scriptsize}

\end{document}